\documentclass[preprint,suppldata]{interact}

\usepackage{epstopdf}% To incorporate .eps illustrations using PDFLaTeX, etc.
\usepackage[caption=false]{subfig}% Support for small, `sub' figures and tables

\usepackage{hyperref}
\usepackage{apacite}
\usepackage{rotating}
\usepackage{xcolor}

\usepackage[longnamesfirst,sort]{natbib}% Citation support using natbib.sty
\bibpunct[, ]{(}{)}{;}{a}{,}{,}% Citation support using natbib.sty
\renewcommand\bibfont{\fontsize{10}{12}\selectfont}% To set the list of references in 10 point font using natbib.sty

\begin{document}

\articletype{Research Article}

\title{A System Dynamics Analysis of National R\&D Performance Measurement System in Korea}

\author{
\name{Taekho You\textsuperscript{a} and Woo-Sung Jung\textsuperscript{a,b}\thanks{CONTACT Woo-Sung Jung. Email: wsjung@postech.ac.kr}}
\affil{\textsuperscript{a}Department of Industrial and Management Engineering, POSTECH,
Pohang 37673, Korea; 
\textsuperscript{b}Department of Physics, POSTECH, Pohang 37673, Korea}
}
\maketitle
	
\begin{abstract}
Peer review is one of useful and powerful performance measurement process. In Korea, it needs to increase quality of R\&D performance, but bibliometric evaluation and lack of peers have opposite effect. We used system dynamics to describe Korean R\&D performance measurement system and ways to increase performance quality. To meet a desired R\&D performance quality, increasing fairness and quality of evaluation is needed. Size of peer pool decreased because of the specialization of R\&D projects and the {\it Sangpi} process both, and it is critical to acquire both fairness and quality. Also, shortening evaluation period affect to R\&D performance quality, by causing workloads increase, limiting long-term and innovative R\&D projects, and decreasing evaluation quality. Previous evaluation policies do a role like micro-controlling the R\&D's activities, but increasing the size of peer pool and changing evaluation period would make a change to quality and fairness of evaluation.

\end{abstract}


\section{Introduction}

Nations invest huge amount of taxes to the R\&D projects for science and technology development. It is important to measure the project's performance, but it is a difficult work. Measurement criteria, increasing number of R\&D projects, science and technology development itself, conflict of interest, and correlation between those phenomena make it more complex. Peer review is the professional tool for evaluating scientific work. Peer review is effective to evaluate scientific work qualitatively because it relies on professional peer's decision. It is used widely such as prizing, performance measurement, and evaluating research group or institution\citep{Hemlin:1996ct}.

The Korean government decided to enhance bibliometric evaluation since 2006. For measuring R\&D performance and managing it the bibliometric indices are used such as SCI impact factor, number of citations, number of papers, and patents. And researcher set their annual goal using these indices. After 2006, R\&D performance has grown remarkably, especially bibliometric indices increased about 30\% annually. However, the bibliometric evaluation is criticized that the qualitative performance is not grown with quantitative one. The government has been trying to improve qualitative performance of R\&D activity, from diverging R\&D investment strategy to managing researchers. Changing R\&D performance measurement system is also one of these improvement policies. It is now on debating how to change the system, for example, improving peer review process and abolishing {\it Sangpi} process has been discussed for a long time. In this paper, we focus on describing and analyzing R\&D performance measurement system and issues with increasing qualitative performance.

We use system dynamics to represent and analyze issues of R\&D performance measurement system in Korea. System dynamics is a useful way to describe issues of a complex system. System dynamics describes the system structure using subsystem diagram, causal loop diagram or stock and flow diagram with feedback loops. It gives insight to the dynamic behavior of the issue that occurs in the system\citep{Forester:1961uw, Sterman:2000hp}. Major mechanisms are represented to feedback loops that are reinforced or counteracted with positive or negative loops, respectively. The changes of R\&D performance measurement system will accompany side effects. Therefore, minimizing these effects is necessary to decision making. To do this, it is important to investigate cause and effect relations in the system. System dynamics offers a chance to understand dynamics of R\&D performance measurement system and political decision to the system \citep{Rodrigues:1996ho}.

Rest of this paper is organized as following: in section 2, literature review of peer review in R\&D performance measurement system is introduced, because peer review is one of powerful and widely-used process to measure performance quality. In section 3, a system dynamics approach of R\&D performance measurement system in Korea will be introduced. Issues of the system and causal loop diagram about them will be described here. Finally, we summarize results of the system dynamics analysis and give insight to the system.

\section{Peer review}

\subsection{Bias and conflict of interest in peer review}

% Measuring R\&D performance is important both to manage R\&D project and effective use of funds\citep{Callier:2015vw}. To measuring it both bibliometric evaluation and peer review are widely used. Bibliometric evaluation is fast and easy evaluation, but it is carefully used for comparing performance between two different R\&D projects. More discussion about bibliometric evaluation's comparing is referred to\citet{}.

For evaluation, peer review is a powerful and effective process because it is determined by professional peers. The problem of peer review, however, is also that it is highly rely on professional peers. It is not easy to object to a decision although peers are independent\citep{Roy:1985fb, Bornmann:2011uk}, so peer selection is important to make a fair evaluation. Even though all know it, previous works show that bias exist such as gender, nationality, prestige, and conservatism. For example, \citet{Wenneras:1997ju} and \citet{Sandstrom:2008be} shows that female and young researchers should make more efforts to get funds in Swedish Medical Research Council (MFR). While CV, bibliography and a research proposal are reviewed by a committee without conflict of interest, and bibliometric indices has no difference between two, men has higher success rate. \citet{Wenneras:1997ju} notes that peers who have relationship give higher score than others, and 'male gender' tend to have this bonus more than women.

In the case of interdisciplinary research, peer review is more complex. Different assessment by different peers for their own research interests, difficulty in judgement, and policy make peers do biased evaluation\citep{Langfeldt:2006hn,Laudel:2006by,Pautasso:2010jm}. Usually interdisciplinary research has more participants, participants are international, or the project are top-down for governmental policy. Bias driven peer selection is easier and decision not related to project quality can be possible.

Conflict of interest is another aspect of the disturbing evaluation. While regulation for conflict of interest, the the evaluation is performed not related to research performance itself. For example, peers often have a different view about the research topic and question. Or they often lie in opposite situation like major competitor. Most of the case peers are needed to be objective, but conflict of interest makes them difficult to be objective.

These works imply two thing. First, peer's relationship is critical factor to affecting evaluation results. If any relation exist peers suffered to give a bonus\citep{Martinson:2005co,Wenneras:1997ju}, because they know about the work more than others or even they already grouped together. Second is that even though independent peers are selected for the evaluation, bias that exist in peer's mind affects to performance evaluation. It is hard to find and classify it, thus hoping for peers to acting fairly\citep{Martinson:2005co}.

The reason of bias in peer review can be summarized as following: peer individual, peer-researcher relation, and inter-peer relation. Variations to avoid bias are suggested, and we refer \citet{Lee:2013ft} as they give a summary for bias of academic peer review. They also summarize types of peer review applied, such as double-blind peer review or open peer review. The fundamental of peer review is that achieving both fairness and its quality, simply said. For example, double-blind peer review achieves it by hiding information about peers and authors and open peer review try opposite way by open information to the both. They have weakness, so deciding which type is proper one is a heuristic problem.

\subsection{Real-world problems of peer review}

While peer review is criticized by its reliability, validity and bias as other studies show, it is useful tool for the alternative of bibliometric evaluation. \citet{Opthof:2010iz} and \citet{Bertocchi:2015gs} suggest a correlation of evaluation results between bibliometric evaluation and peer review. \citet{Opthof:2010iz} studied that reports of CWTS and leading scientists's publication have same ranking structure of Netherland's Medical Physics research. While debates exist about this correlation after the study, such as \citet{vanRaan:2010cn}, \citet{Bornmann:2010eg}, and \citet{Waltman:2011if}, we can agree that they are not independently evaluated. \citet{Bertocchi:2015gs} studied performance difference between peer review of Economics, History, Management and Statistics in Italy and its paper's journal indices. Even though these research sub-areas have different bibliometric indices, the bibliometric ranking and peer review result give same classification results. The difference between two result have no statistically significant, whether peer review is double-blinded or not. While these studies show evaluation results are statistically no different, however, we treat them as different one because the two evaluations has different effect in the system.

\begin{sidewaysfigure}
\includegraphics[width=\columnwidth]{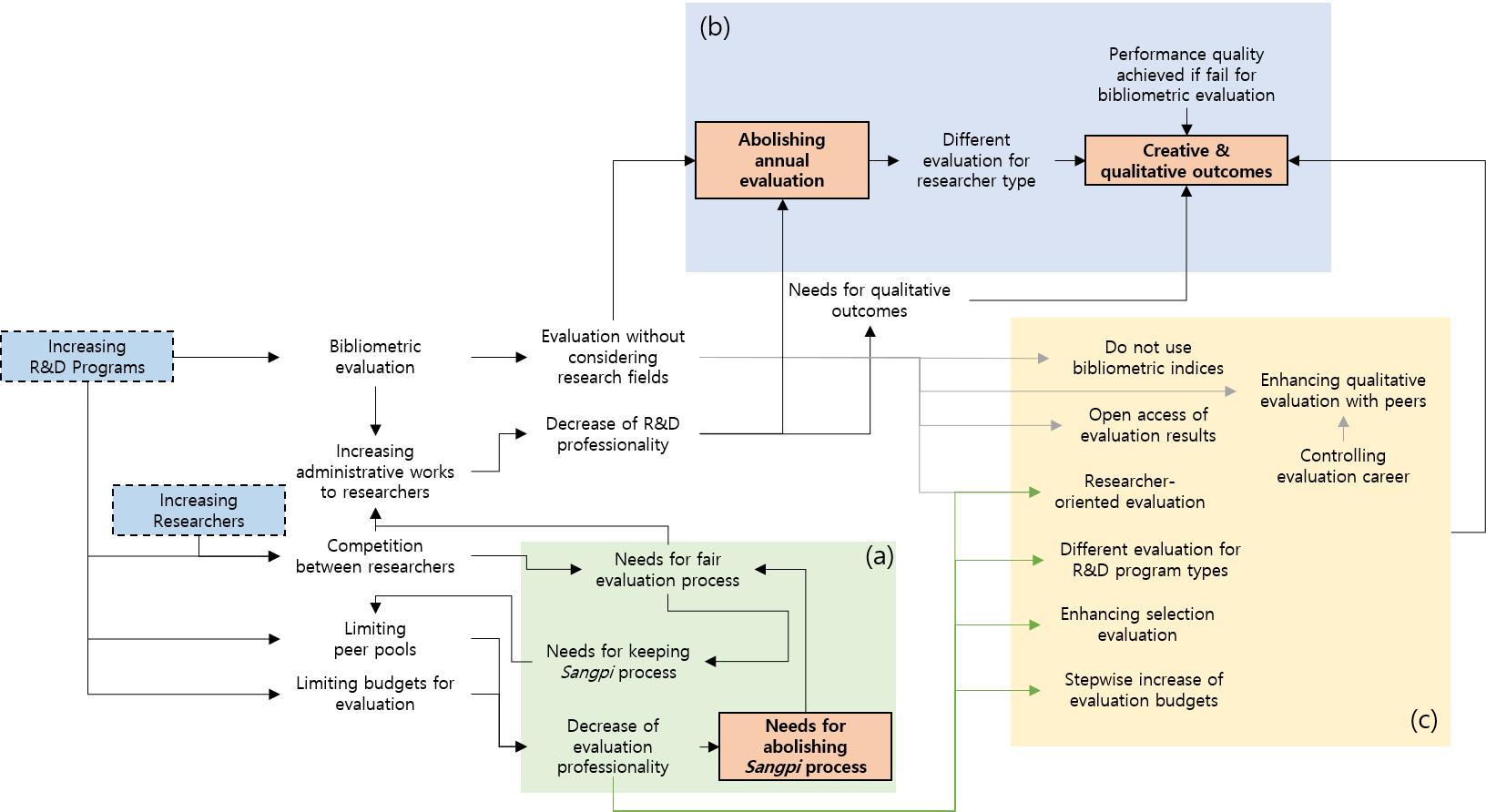}
\caption{Subsystem diagram of the R\&D performance measurement system issues}
\label{fig:sd}
\end{sidewaysfigure}

In Korea, a peer selection policy to prevent bias and acquiring fairness is {\it Sangpi} process. The {\it Sangpi} process is that exclude possible peers who are related to the R\&D participants at the peer selection step. Before the selecting peer, an ego-centric network about R\&D participants were made; if peer-peer or peer-researcher has any relation, they are linked. After that, peers within a distance are excluded. From doing this, a peer is excluded if peer is in same university or institution, they graduate same school, or they participated in the same project before with one of R\&D participants. Controlling size of distance peers who are professional and have no relation can be selected, but practical problem arise. As R\&D fields are highly specialized network size in a field is decreased. If distance increases, peers selected are too far from the project so it could not get reliable evaluation results. If distance decreases, peers selected are may have relation to R\&D project participants, so fairness problem arise. The issue about {\it Sangpi} process is to abolish and keeping it, described in Fig.~\ref{fig:sd}(a).

\section{Analysis of R\&D Performance Measurementment System in Korea}

\subsection{System and Issues Description}

As a start we use subsystem diagram to summarize practical issues. Note that to summarize we used reports in two Korean Institute, the Science and Technology Policy Institute (STEPI) and Korea Institute of S\&T Evaluation and Planning (KISTEP), written in Korean. Fig.~\ref{fig:sd} is a subsystem diagram of the National evaluation system in Korea. As an international trend to increase R\&D funds, the Korean government also increases number of R\&D programs. Also they decided to enhance bibliometric evaluation since 2006, which determine a R\&D work's achievement by number of papers, impact factor, number of patents, etc. Bibliometric evaluation system is effective to control all R\&D programs, because scoring R\&D performance by numerical indices is easy to compare them. However, bibliometric evaluation gives two effect to the R\&D performance evaluation system. First, they used evaluation results to comparing R\&Ds without considering their research fields. Bibliometric indices are different values in research fields and comparing between two other works is dangerous. Second, researchers take more administrative workload. At the project selection step, researchers propose their annual goals such as how much they public papers or patents each year. They should summarize and present their annual work more and more because competition between researchers increases and evaluation quality decrease.

Fig.~\ref{fig:sd}(a) shows an issue about {\it Sangpi} process. As we mentioned, the size of peer pool is controlled by the distance, and maximum number of peers are decreased when R\&D project is more specialized. The limitation of peer pool and budget, the size of peer pool decreased and achieving evaluation quality and fairness is hard. So there has voiced of abolishing {\it Sangpi} process. But if abolishing, unfair peers would be selected more than before, so the problem would not be solved clearly. System dynamics analysis of the process is described more in Fig.~\ref{fig:peerpool}.

R\&Ds becomes interdisciplinary, and they are needed to produce more qualitative performance. There has voiced that evaluation system blocks for researchers to produce by creative and qualitative performance. The issue is about the annual evaluation, shown in Fig.~\ref{fig:sd}(b). The annual evaluation has been adopted to manage ongoing R\&D projects. However, increasing the number and size of reports make it hard to control. It eventually decrease time to focusing on R\&D work, leading performance quality decreases. Some of creative performance cannot be produced annually and annual evaluation is the opposite evaluation process to doing these R\&Ds. Fig.~\ref{fig:challengeable} described more about the issue.

As a summary, we show two issues in Korean R\&D performance evaluation system, {\it Sangpi} process and annual evaluation. The system dynamics analysis for the system will be given in Section~\ref{sec:SD}, but additional suggestions are given in Fig.~\ref{fig:sd}(c). For example, increasing budget can contribute to solving such as lack of peer pool size and quality increase by give plenty of time with incentive to peers. However, they are a part of solving problems and do not affect dynamics of the evaluation system to be changed.

\subsection{System Dynamics Analysis of the System}
\label{sec:SD}

\begin{figure}
\includegraphics{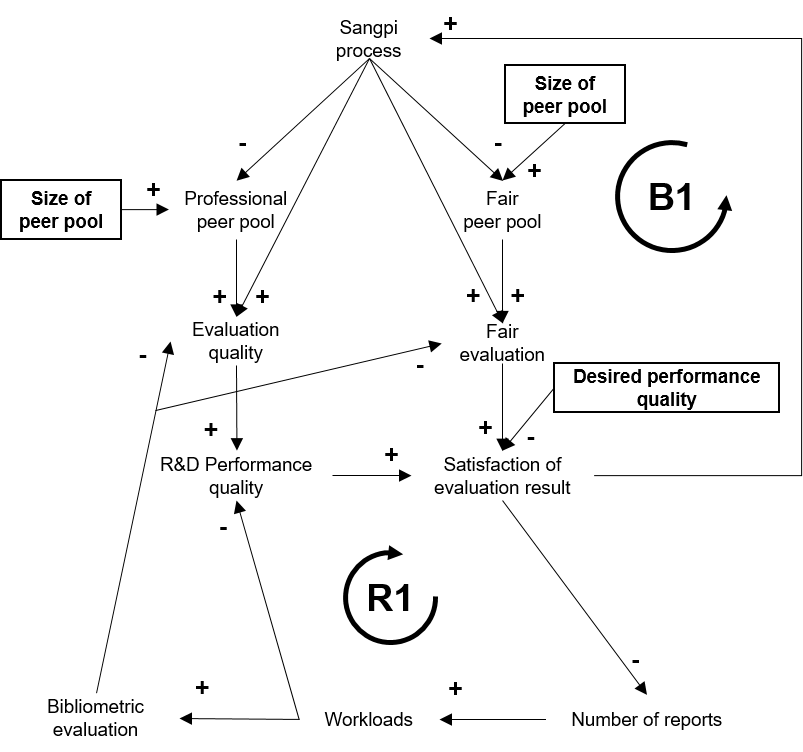}
\caption{Cycles describing the qualitative R\&D evaluation}
\label{fig:peerpool}
\end{figure}

A causal loop diagram of the {\it Sangpi} process and the annual report about R\&D work is described in Fig.~\ref{fig:peerpool}. We draw two loops, one is a negative feedback loop for the {\it Sangpi} process and the other is a positive feedback loop for the annual report. Both loops affect to evaluation quality and fairness, and R\&D performance quality and satisfaction for the evaluation results is controlled by two policies. Note that there will be other uncountable factors for evaluation. For example, an incentive to successful R\&D project\citep{Squazzoni:2013er} will give motivation to the researchers, but we do not handle these factors to focus on issues.

The role of {\it Sangpi} process is to exclude peers who are not professional and have any relation to the R\&D project. With the {\it Sangpi} process reliable evaluation results with no bias can be achieved, but this process also decreases the number of possible peers. If size of peer pool is enough it causes no problems, but if not it will be hard to select professional and fair peers.

So the loop B1 depends on size of peer pool. As Korean ministry restricts peers to Korean researchers, there is limited peer pool in the system. Accompany with that, R\&D projects are highly specialized, finding professional peers without relation becomes a hard work. Thus quality of R\&D evaluation decrease, and the unsatisfaction of evaluation results leads unsatisfaction for competitive researchers. They would insist that it is needed to repeal {\it Sangpi} process to monitor R\&D project strongly.

To control the R\&D projects, Korean government ministry requires more reports to R\&D researchers, such as the annual report or increase the detail of each report. It would lead to increase of workloads to both R\&D researchers and the ministry. More work would decrease project work time, which would cause decrease of performance quality. And to handle more reports, the ministry evaluates performance by bibliometric indicators. The procedure is described in the loop R1. Since it is a positive feedback loop, it expects that exponential growth of workloads and performance decrease, but the limited budget and time controls this exponential growth while we do not draw in the Fig.~\ref{fig:peerpool}.

Repealing {\it Sangpi} process is not the solution since the process is essential to control peer evaluation process. As we mentioned, size of peer pool is an important factor to {\it Sangpi} process. Regulation of peer pool should be expanded, including foreign researchers and experts in industrial area. For example, for the commecialized R\&D programs patent attorney or lawyer can be appropriate to evaluate R\&D performance. As \citet{Marsh:2008ic} noted that external peers have rare experience for evaluation and training them is another important issue. For doing this, more funds to the peer review process and selecting appropriate peers by its type of R\&D program will be needed. The more the size of peer pool increases, the more professional and fair evaluation are achieved.

\begin{figure}
\includegraphics[width=\columnwidth]{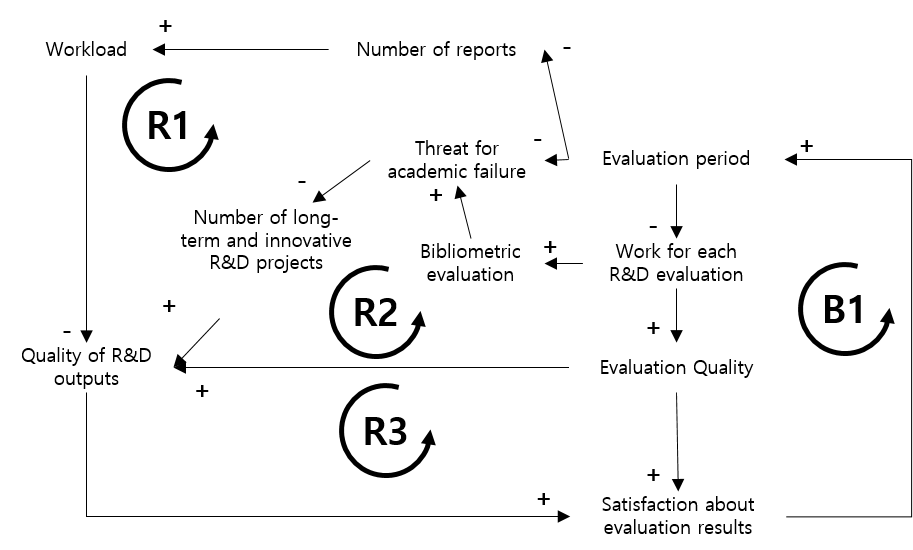}
\caption{Cycle describing the challengeable R\&D works}
\label{fig:challengeable}
\end{figure}

To focus on another issue of annual evaluation, we draw another causal loop diagram in Fig.~\ref{fig:challengeable}. We start by evaluation period, which is the duration between two sequential evaluation in a R\&D project. This value can be varied from each day to whole R\&D project period that means only projection selection and final evaluation exist. However, it is not controlled by other variables, but decided by the policy in each country's evaluation strategy and for considering R\&D characteristics. We draw it as a balancing loop B1 in Fig.~\ref{fig:challengeable}, because the evaluation period cannot be changed easily, like a stable state, and evaluation period affect to itself eventually.

Evaluation period affect to quality of R\&D outputs by three ways as we draw these reinforcing loops in Fig.~\ref{fig:challengeable}. These are about research works, researcher's career, and evaluation quality itself. First, as evaluation period are shortened, R\&D participants should produce more reports. Making each report is time-consuming work, so it would limit time to focus on R\&D activities relative to long evaluation period. Second, researchers should produce their research outcomes in a short time if evaluation period becomes shorten. Moreover, if evaluation is done by bibliometric indices, they need number of papers, patents, or any achievements in the period. In this case, the threats make researchers to avoid some of disciplines and so-called innovative R\&D project because they need time to produce outcomes and have higher probability to failure. The last loop is about evaluation quality. Demanding work for R\&D evaluation increase if evaluation period is shortened. We can easily infer that increasing work with limited resources means decrease of evaluation quality. If evaluation process does not work normally, the probability for abusing would increase.

As we mentioned, evaluation period has been optimized and decided by the policy. Changing evaluation period will give tremendous effect to the performance measurement system, because it does not mean the only changes of period itself but would imply changes of R\&D management policy, strategy, and researcher's career management. Nevertheless, we suggest increasing period to ensure that researchers try innovative R\&D works. To do that, qualitative evaluation is also needed to determine their work's value if they failed. Fig.~\ref{fig:sd}(c) would be good strategies to concrete the idea.

\section{Conclusion}

In this work we analyzed Korean R\&D performance measurement system, showing that bibliometric evaluation increase workload and negative effect for researchers to try creative works from the positive feedback loop. It is the "truth deficit" of the evaluation. For the trend of increasing interdisciplinary research and demand for creative performance, enhancing bibliometric evaluation is not proper. And Korean specialized process, named {\it Sangpi}, exists to limit the size of peer pool for increasing fairness and evaluation quality, but it also has negative effect to them. From this we suggest that repealing the process is not the solution so increasing the size of peer pool with external peers will be needed.

We analyze issues by system dynamics approach, and propose these political suggestions as a possible policy. The system dynamics approach provides information for project management. The results would give insight to make a political decision, but it needs to analyze the decision's impact quantitatively. However, policy test is not a simple work, even starting by measuring quantities of each variables is hard to achieve. As it is not an easy work to apply these policy suggestion in the real world, a more precise modeling of the system and simulation would be proceded.

Peer review is becoming more important process to measuring performance qualitatively, both project management and academic area. We proposed a system dynamical approach to the R\&D performance evaluation in Korea, it will be benefit to any complex system of evaluation. The approach can be expanded to analyze various administrative systems. It is needed to improve with political decision making since most decisions of administrative systems are irreversible. Although simulation of such system is hard to establish, mapping causal relation considering stakeholders and their issues is helpful to decision making. In academic area, bias like nationality and conservatism is one good expansion to apply the analysis.

\section*{Acknowledgement}
This work was supported by Basic Science Research Program through the National Research Foundation of Korea (NRF) funded by the Ministry of Education (2016R1D1A1B03932590).

\bibliographystyle{plainnat}
\bibliography{AJTI}

\end{document}